\documentclass[3p,times]{elsarticle}

\usepackage{ecrc}

\volume{00}

\firstpage{1}

\journalname{Physics Procedia}

\runauth{A. Renshaw, for the Super-Kamiokande Collaboration}

\jid{procs}

\jnltitlelogo{Physics Procedia}

\usepackage{amssymb}
\usepackage{amsthm}
\usepackage{amsmath}
\usepackage{graphicx}
\usepackage{dcolumn}
\usepackage{epstopdf}
\usepackage{caption}
\usepackage{subcaption}

\usepackage{lineno}

\usepackage[figuresright]{rotating}

\begin{document}

\begin{frontmatter}


\dochead{13th International Conference on Topics in Astroparticle and Underground Physics}

\title{Solar Neutrino Results from Super-Kamiokande}

\author{Andrew Renshaw, for the Super-Kamiokande Collaboration}

\address{Department of Physics and Astronomy, University of California, Irvine\\3117 Frederick Reines Hall, Irvine, California 92697, USA}

\ead{arenshaw@uci.edu}

\begin{abstract}
Super-Kamiokande-IV (SK-IV) data taking began in September of 2008, after upgrading 
the electronics and data acquisition system.  Due to these upgrades and
improvements to water system dynamics, calibration and 
analysis techniques, a solar neutrino signal could be extracted at 
recoil electron kinetic energies as low as 3.5 MeV.  When the SK-IV data
is combined with the previous three SK phases, the SK extracted 
solar neutrino flux is found to be 
$[2.37\pm0.015\mbox{(stat.)}\pm0.04\mbox{(syst.)}]\times10^6$/(cm$^{2}$sec).  
The combination of the SK recoil electron energy spectra slightly favors distortions 
due to a changing electron flavor content.  Such distortions are predicted when
assuming standard solar neutrino oscillation solutions. An extended 
maximum likelihood fit to the amplitude of the expected solar zenith
angle variation of the neutrino-electron elastic scattering rate
results in a day-night asymmetry of 
$[-3.2\pm1.1$(stat.)$\pm0.5$(syst.)]$\%$.
A solar neutrino global oscillation analysis including all current solar neutrino
data, as well as KamLAND reactor antineutrino data, measures the solar mixing angle as
$\sin^2\theta_{12}=0.305\pm0.013$, the solar neutrino mass squared splitting as 
$\Delta m^2_{21}=7.49^{+0.19}_{-0.17}\times10^{-5}$eV$^2$ and 
$\sin^2\theta_{13}=0.026^{+0.017}_{-0.012}$.
\end{abstract}


\end{frontmatter}


\section{Introduction}
\label{intro}
Solar neutrino flux measurements from the Super-Kamiokande (SK)~\cite{sk1} and
the Sudbury Neutrino Observatory (SNO)~\cite{sno1} experiments provided direct
evidence that the deficit of solar neutrinos observed by the Homestake~\cite{homestake}
and other solar neutrino experiments is the result of solar neutrino flavor conversion.
While this solar neutrino flavor conversion is well described by neutrino
oscillations (in particular oscillation parameters extracted using solar
neutrinos agree with those extracted using reactor antineutrinos
\cite{kamland}), there is still no direct evidence for this to be so.  It is
possible that the flavor conversion is driven by some other mechanism.
However, based on the current model and parameters of solar neutrino
oscillations, there are two testable signatures available for the SK
experiment to look for.  The first is the observation and precision
measurement of the expected Mikheyev-Smirnov-Wolfenstain (MSW)~\cite{msw}
resonance curve.  Based on the 
current best-fit oscillation parameters extracted using both solar neutrino and
reactor
antineutrino data, there is an expected characteristic energy dependence of the
flavor conversion.  Higher energy solar neutrinos, such as $^8$B and $hep$
neutrinos, undergo complete resonant conversion within the Sun, while lower
energy solar neutrino, such as $pp$, $^7$Be, $pep$, CNO and the lowest energy
$^8$B neutrinos, only suffer from vacuum oscillations.  After averaging
the vacuum oscillations due to energy resolution, the survival probability for
low energy electron flavor solar neutrinos must exceed $50\%$, while the 
resonant
conversion of the higher energy solar neutrinos within the Sun leads to the
currently observed survival probability of about $30\%$.  The transition
between the vacuum dominated and solar resonance dominated oscillations should
occur near three MeV, making $^8$B solar neutrinos the best choice when
searching for the transition point within the energy spectrum.  

The second solar neutrino oscillation signature comes from the effect of the
terrestrial matter density.  This effect can be tested directly by comparing
the rate of solar neutrino interactions during the daytime to the rate during
the nighttime, when the solar neutrinos have passed through the Earth.  After
being resonantly converted into the second mass eigenstate within the Sun, the neutrinos
which then pass through the Earth will generally have an enhanced electron
neutrino flavor content.  This will lead to an excess in the electron
elastic scattering rate during the nighttime, and hence a negative
``day-night asymmetry'' $A_{\text{\tiny DN}}=(r_{\text{\tiny D}}-r_{\text{\tiny N}})/r_{\mbox{\tiny ave}}$, where $r_{\text{\tiny D}}$
($r_{\text{\tiny N}}$) is the average daytime (nighttime) rate and
$r_{\mbox{\tiny ave}}=\frac{1}{2}(r_{\text{\tiny D}}+r_{\text{\tiny N}})$
is the average rate.  SK observes a wide range of $^8$B solar neutrinos, making
it a prime detector to search for both of the solar neutrino oscillation
signatures.

The most recent solar neutrino results from the SK experiment have been
presented.  This includes the latest flux measurement from the fourth phase
of SK (SK-IV), energy spectrum and day-night asymmetry analyses using all SK
data and oscillation analyses using SK data only and then SK data plus all
other relevant data (other solar neutrino and reactor anti-neutrino data).
Complete details of these analyses can be found in~\cite{sk4,skall_dn}.

\section{Super-Kamiokande IV Improvements}
\label{sk4}
Super-Kamiokande is a 40 m diameter, 40 m tall right cylindrical stainless
steel tank filled with 50 kton of ultra-pure water, located in Kamioka, Japan.
The detector is optically separated into 2 distinct volumes, a 32 kton inner
detector (ID) and a 2 m active veto outer detector (OD) surrounding the ID.
The structure used to divide the two volumes houses an array of 11,129 50 cm
photo-multiplier tubes (PMTs) facing the ID and 1,885 20 cm PMTs facing the OD.
The detector itself is currently in the same configuration as during the SK-III
phase~\cite{sk3}, however improvements to the data acquisition system (DAQ) marked
the end of SK-III and the beginning of SK-IV.

SK-IV began data taking in September of 2008, after having all of its front-end
electronics upgraded.  The new boards, called QBEEs (QTC Based Electronics
with Ethernet Readout)~\cite{qbee}, allowed for the development of a new online
DAQ.  The essential components of the QBEEs, used for
the analog signal processing and digitization, are the QTC (high-speed
Charge-to-Time Converter) ASICs, which achieve very high speed signal
processing and allow the readout of every hit of every PMT.  The resulting hit
PMT information is sent to online computers which scan the data and use a
software trigger to select time coincidences within 200 nsec, in order to
pick out events.  The software trigger ensures that a high rate of super low
energy events does not impact the efficiency of triggering on high energy events
and allows for flexible event time windows.  The energy threshold using this 
software trigger is only limited by the speed of the online computers, and is
set at 3.5 MeV recoil electron kinetic energy, the lowest of all SK phases. 
The triggering efficiency of SK-IV events is better than
$99\%$ at 4.0 MeV and $\sim84\%$ between 3.5 and 4.0 MeV.

Because of the large size of SK, it is necessary to continuously recirculate
the water to maintain optimal water clarity.  This is done by extracting water
from the top of the detector,
sending it through a water purification system and then re-injecting it into
the bottom of the detector.  If the temperature of the water being injected
into the bottom of the tank is not closely matched to that of the rest of the
detector, convection will occur within the tank.  This allows
radioactive radon (Rn) gas, which is most commonly produced near the edge of
the detector by decays from the U/Th chain, to make its way into the central
region of the detector.  Radioactivity coming from the decay products of
$^{222}$Rn, most commonly $^{214}$Bi, can mimic the recoil electron signal
coming from the elastic scattering of a solar neutrino.  In January of 2010,
a new automated temperature control system was installed to control the
temperature of the water being injected into the detector at the $\pm$0.01 K
level.  By controlling the supply water temperature and the rate at which
water is extracted and injected to different places in the detector,
convection within the tank has been kept to a minimum and the background level
in the central region has become significantly lower, compared to SK-III.

Besides the above hardware improvements to the detector, a new analysis method
was introduced to separate background and signal events.  Even at the low
energies of solar neutrinos, it is still possible to use the PMT hit patterns
to reconstruct the amount of multiple Coulomb scattering a recoil electron will
incur.  As the energy of the recoil electron is decreased, the amount of
multiple scattering the electron will incur increases, thus leading to a
more isotropic PMT hit pattern.  The majority of the low energy background in
SK is believed to be coming from the $\beta$-decay of $^{214}$Bi, which has
an endpoint kinetic energy of $\sim2.8$ MeV.  With the low energy threshold of
SK-IV set at 3.5 MeV, the only way these
lower energy $\beta$-decays contaminate the solar neutrino data set is due to
Poisson fluctuations of the number of reconstructed photons, resulting in a larger
reconstructed energy.  However, despite these events fluctuating up in energy, they
should still multiple scatter as electrons with kinetic energy less than 2.8
MeV.  These $\beta$-decays should therefore undergo more multiple scattering
than the solar neutrino interactions.
SK-IV has introduced a new multiple Coulomb scattering goodness (MSG) variable,
described in detail in~\cite{sk4}, allowing data events to be broken into
sub-samples based on the amount of multiple scattering, before the solar
neutrino signal is extracted.

\section{Detector Performance}
\label{detector_performance}
The methods used for the vertex, direction and energy reconstructions are the
same as those used for SK-III~\cite{sk3}.  There is a very slight improvement in
the vertex resolution during the SK-IV phase ($\sim50$ cm at 9.5 MeV), compared 
to SK-III, the result of
improved timing resolution and timing residual agreement between data and MC
simulated events coming from the upgraded front-end electronics.  The angular
and energy resolutions are nearly identical to the SK-III phase, $\sim25^{\circ}$ 
and $\sim14\%$ for 9.5 MeV electrons, respectively.  The absolute energy scale
is determined with a small electron linear accelerator (LINAC), which 
injects single monoenergetic
electrons into the SK tank, in the downward direction, with energies between
4.2 and 18.5 MeV. More details are described in~\cite{linac}.  The energy of
the LINAC electrons are precisely measured by
a germanium (Ge) detector.  The directional and position dependence of the
energy scale is further check using a deuterium-tritium (DT) fusion neutron
generator~\cite{dt}.  The total error on the absolute energy scale
resulting from these calibrations is found to be $0.54\%$, similar to
the SK-III value of $0.53\%$.  

The water transparency (WT) in the MC simulation is defined using
absorption and scattering coefficients
as a function of wavelength (see~\cite{sk4calib} for details).  The
dominant contribution to the variation of the WT is a variation in the
absorption length.  The scattering coefficients are taken as
constants, while the absorption coefficient
is both time and position dependent.  The time variation of the
absorption coefficient is checked using the light attenuation of Cherenkov light from decay
electrons, resulting from cosmic-ray $\mu$'s.
The position dependence of the absorption coefficient arises from draining
water from the top of the detector and re-injecting it into the bottom as it
is continuously recirculated.  Due to the precise control of the input water
temperature, the convection inside the tank is minimized everywhere but the
bottom, below $z=-11$ m.  Due to a small amount of convection in the bottom
of the tank and a constant rising temperature above, the absorption
coefficient is modeled as a constant below $z=-11$ m and with a linear function
above this height.  This ``top-bottom'' asymmetry of the WT is determined
by studying the distribution of hits coming from a Ni-Cf gamma-ray source
(see~\cite{sk4calib}) in
the ``top'', ``bottom'' and ``barrel'' regions of the detector.  It is found
that the hit rate of the top region of the detector is $3\sim5\%$ lower than
that of the bottom region.  The time dependence of this top-bottom asymmetry
is monitored using the same Ni calibration, as well as an auto-xenon
calibration~\cite{sk4calib}.  The introduction of this time dependent absorption coefficient
has much reduced the systematic uncertainty resulting from the directional
dependence of the energy scale, especially useful for the solar neutrino
day-night asymmetry analysis.

\section{Data Reduction}
\label{reduction}
The majority of the analysis cuts are the same as used for the SK-III
phase~\cite{sk3}, however, in order to optimize the significance
$(S/\sqrt{BG})$, the applied energy regions have slightly changed and a new
tight fiducial volume cut is applied. Events between 4.5 and 5.0 MeV
are cut if the radius squared $r^2$ is larger than 180 m$^2$ or the height
$z$ is less than -7.5 m.  Below 4.5 MeV, events are cut if they do not satisfy
\begin{equation}
\frac{r^2}{\mbox{m}}+\frac{150}{11.75^4}\times\left|\frac{z}{\mbox{m}}-4.25\right|^4 \le 150,
\end{equation}
with the coordinates given in meters.  The remaining efficiency above 6.0 MeV
is almost identical to SK-III, while for 5.0 to 6.0 MeV, SK-IV is better
than SK-III.  This is caused by removing the second vertex cut and making
a looser ambient event cut.  Using the new tight fiducial volume cut and a
tighter ambient event cut for 3.5 to 5.0 MeV gives a lower selection
efficiency, however, in exchange the background level has been much reduced.

\section{Data Analysis}
\label{data_analysis}
\subsection{Total Flux}
\label{flux}

The start of SK-IV physics data taking occurred on October 6th, 2008.  The
results presented include data through the end of December 31st, 2012, a total
of 1306.3 live days.  As opposed to SK-III, which had different livetimes for
the different low energy threshold periods, SK-IV took all data with the same
low energy threshold of 3.5 MeV recoil electron kinetic energy.  SK
observes all flavors of solar neutrinos through the process of
neutrino-electron elastic scattering, however, the total cross section for
electron flavor neutrinos is roughly
six times larger than that of the muon or tau neutrinos.  This comes
from the inclusion of both the charged-current (CC) and neutral-current (NC)
interactions for electron flavor neutrinos, whereas the
muon and tau flavors interact via the NC interaction only, making SK most
sensitive the electron flavor solar neutrinos.

\begin{figure}[t]
\begin{subfigure}{0.49\textwidth}
\vspace*{0.15cm}
 \includegraphics[keepaspectratio=false,height=8.8cm,width=\textwidth,clip]{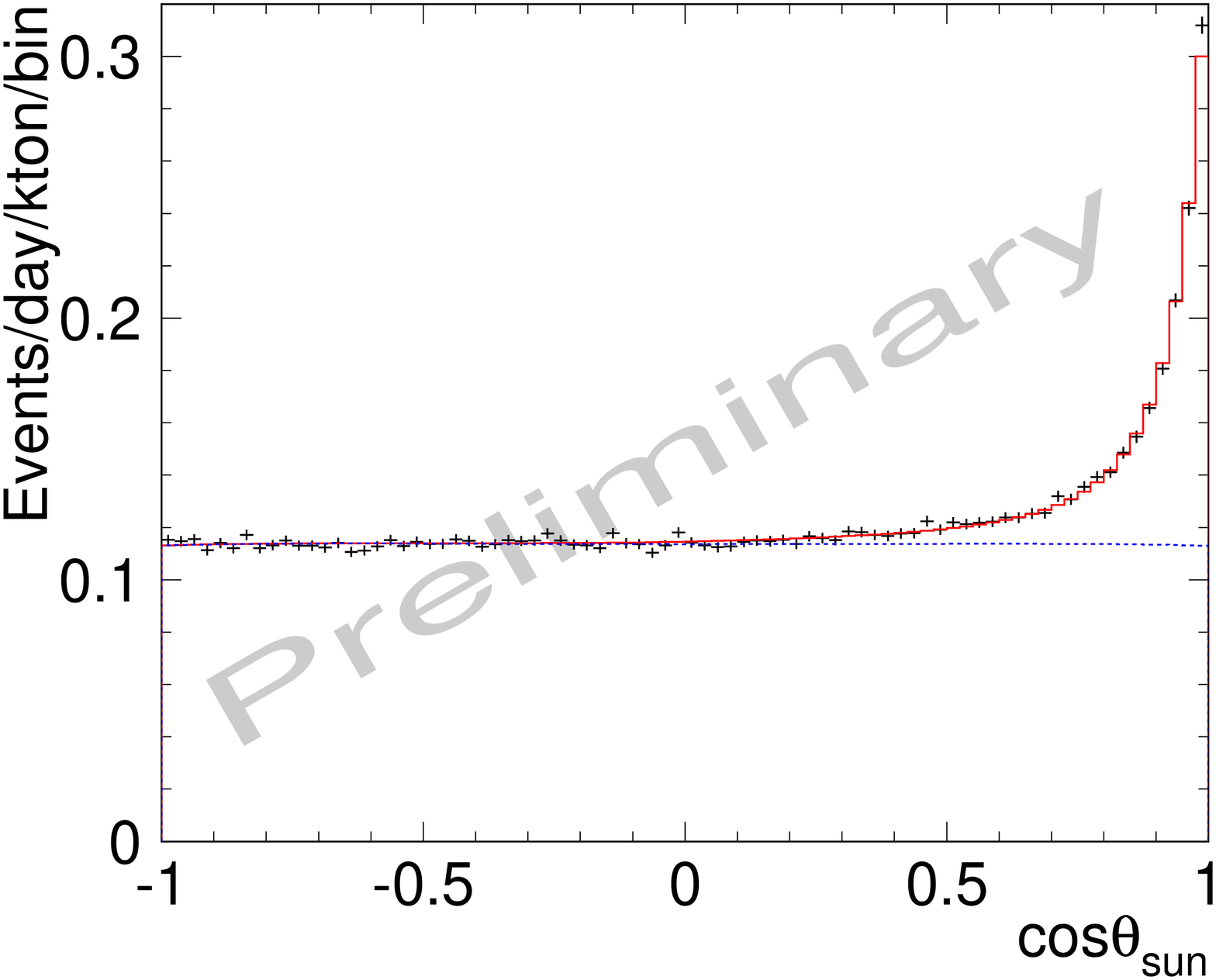}
\end{subfigure}
\begin{subfigure}{0.49\textwidth}
 \includegraphics[width=\textwidth,clip]{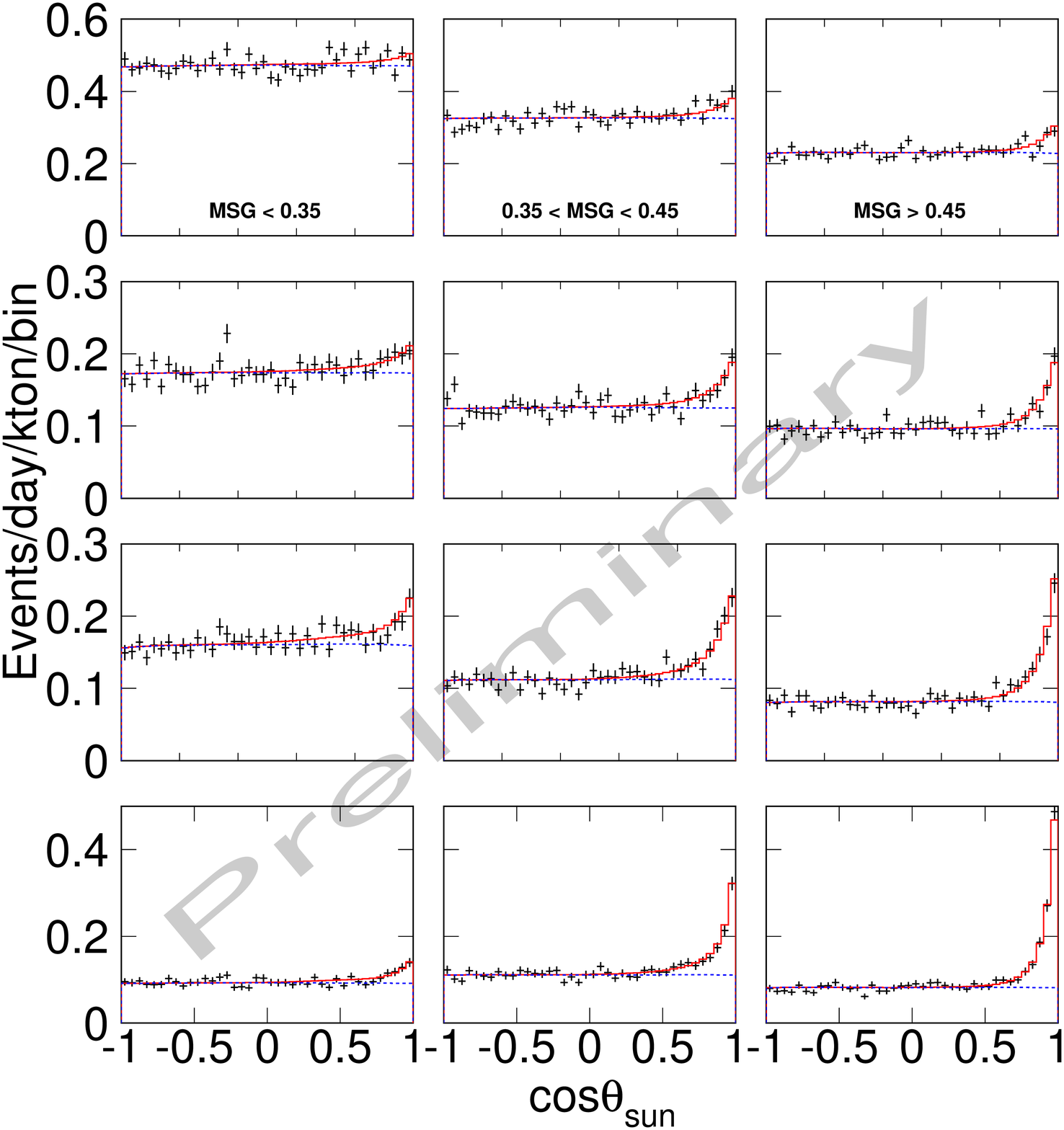}
\end{subfigure}
 \caption{Left: SK-IV solar angle distribution for 3.5 to 19.5 MeV.
$\theta_{\mbox{\tiny sun}}$
is the angle between the incoming neutrino direction and the
reconstructed recoil electron direction. Black points
are data while the blue and red histograms are best fits to the background and
signal plus background, respectively.
Right: Distribution of $\cos\theta_{\text{\tiny sun}}$ for the energy ranges
3.5-4.0 MeV, 4.0-4.5 MeV, 4.5-5.0 MeV and 7.0-7.5 MeV (from top to
bottom), for each MSG bin (left to right).
The colors are the same as the left panel.}
\label{fig:cossun}
\end{figure}

The differential cross section for this interaction, at the
energies of solar neutrinos, is strongly peaked in the direction of the incoming
neutrino.  If $\theta_{\mbox{\tiny sun}}$ is the angle between the incoming solar
neutrino (which is the directional vector from the Sun to the event vertex)
and the reconstructed recoil electron direction, the solar neutrino signal
should peak at $\cos\theta_{\mbox{\tiny sun}}=1$, while background events will be
mostly uniformly distributed.  SK utilizes this by using an extended maximum
likelihood fit between 3.5 and 19.5 MeV recoil electron kinetic energy 
to extract the solar neutrino flux.  The same method is used
for SK-I~\cite{sk1}, SK-II~\cite{sk2} and SK-III~\cite{sk3}.
The left panel of Fig.~\ref{fig:cossun} shows the $\cos\theta_{\mbox{\tiny sun}}$
distribution of the 
SK-IV final data sample (black points), along with the best-fit of the
background (blue) and background plus solar neutrino signal (red).

The systematic uncertainties on the total flux for SK-IV were calculated using
the same methods as for SK-III~\cite{sk3} (see~\cite{sk4} for full systematic
uncertainty details).  The total systematic uncertainty
of the SK-IV flux was found to be $1.7\%$, improved from the $2.2\%$ seen in
SK-III, and the best value among all phases.  The main contributions to the
reduction come from improvements in the uncertainties arising from the
energy-bin uncorrelated uncertainties; the vertex
shift, trigger efficiency and the angular resolution.  There is also a reduction in
the uncertainties associated with the energy scale and resolution, coming from the
addition of the two lowest energy bin, 3.5-4.5 MeV, for the entire period of
SK-IV, compared to SK-III which use a low energy threshold of 6.0 MeV for the
first half of the phase, and 4.5 MeV for the second half.  The installation of the new front-end
electronics has lead to a slightly better timing resolution and agreement
of the timing residuals between data and MC simulated events.  The total number
of solar neutrino events extracted via the extended maximum likelihood fit for
the SK-IV phase is $25,253^{+252}_{-250}(\mbox{stat.})\pm455(\mbox{syst.})$.
This number corresponds to a $^8$B solar neutrino flux of

\begin{align*}
\Phi_{^8\text{B}}(\text{SK-IV})=
[2.36\pm0.02(\text{stat.})\pm0.04(\text{syst.})]\times 10^6 /(\text{cm}^2\text{sec}),
\end{align*}
assuming a pure $\nu_e$ flavor content.  As seen in Table~\ref{tab:flux}, the
flux measurements from each phase of SK agree within the statistical errors.
These four measurements can be combine together to give the total SK-I-IV
combine flux of

\begin{align*}
\Phi_{^8\text{B}}(\text{SK})=
[2.37\pm0.015(\text{stat.})\pm0.04(\text{syst.})]\times 10^6 /(\text{cm}^2\text{sec}).
\end{align*}

\begin{table}[h]
\begin{center}
  \caption{SK measured solar neutrino flux by phase.}
  \begin{tabular}{l c c}
  \hline\hline
           & Energy Threshold & Flux ($\times10^6$/(cm$^2$sec)) \\ \hline
  SK-I      & 4.5 MeV & $2.38\pm0.02\pm0.08$ \\
  SK-II     & 6.5 MeV & $2.41\pm0.05^{+0.16}_{-0.15}$ \\
  SK-III    & 4.5 MeV & $2.40\pm0.04\pm0.05$ \\
  SK-IV    & 3.5 MeV & $2.36\pm0.02\pm0.04$ \\ \hline
  Combined    &  & $2.37\pm0.02\pm0.04$ \\
  \hline\hline
  \end{tabular}
  \label{tab:flux}
\end{center}
\end{table}

\subsection{Solar Neutrino Energy Spectrum}
\label{spectrum}
Solar neutrino flavor oscillations above about 5.0 MeV are dominated by
the solar MSW~\cite{msw} resonance, while low energy solar neutrino flavor
changes are dominated by vacuum oscillations. Since the MSW effect rests
solely on standard weak interactions, it is rather interesting to confront
the expected resonance curve with data. Unfortunately multiple Coulomb
scattering prevents the kinematic reconstruction
of the neutrino energy in neutrino-electron elastic scattering interactions.
However, the energy of the recoiling electron still provides a lower
limit to the neutrino's energy. Thus, the neutrino spectrum is inferred
statistically from the recoil electron spectrum. Moreover, the differential
cross section of $\nu_{\mu,\tau}$'s is not just a factor of about six smaller
than the one for $\nu_e$'s, but also has a softer energy dependence.  In this
way, the observed recoil electron spectrum shape depends both on the flavor
composition and the energy-dependence of the composition of the solar
neutrinos. So even a flat composition of $33\%$ $\nu_e$
and $67\%$ $\nu_{\mu,\tau}$ still distorts the recoil electron spectrum
compared
to one with $100\%$ $\nu_e$. The energy dependence of the day-night effect and
rare $hep$ neutrino interactions (with a higher endpoint than $^8$B $\nu$'s)
also distort the spectrum. To analyze the
spectrum, we simultaneously fit the SK-I, II, III and IV spectra to their
predictions, while varying the $^8$B and $hep$ neutrino fluxes within
uncertainties. The $^8$B flux is constrained to
$[5.25\pm0.20]\times10^6$ /(cm$^2$sec)
and the $hep$ flux to $[2.3\pm2.3]\times10^4$ /(cm$^2$sec)
(motivated by SNO's measurements~\cite{snothreephase,snohep}).

\subsubsection{SK-IV Energy Spectrum}
\label{sk4spec}
\begin{figure}[t]
\begin{subfigure}{0.48\textwidth}
 \begin{center}
 \includegraphics[width=\textwidth]{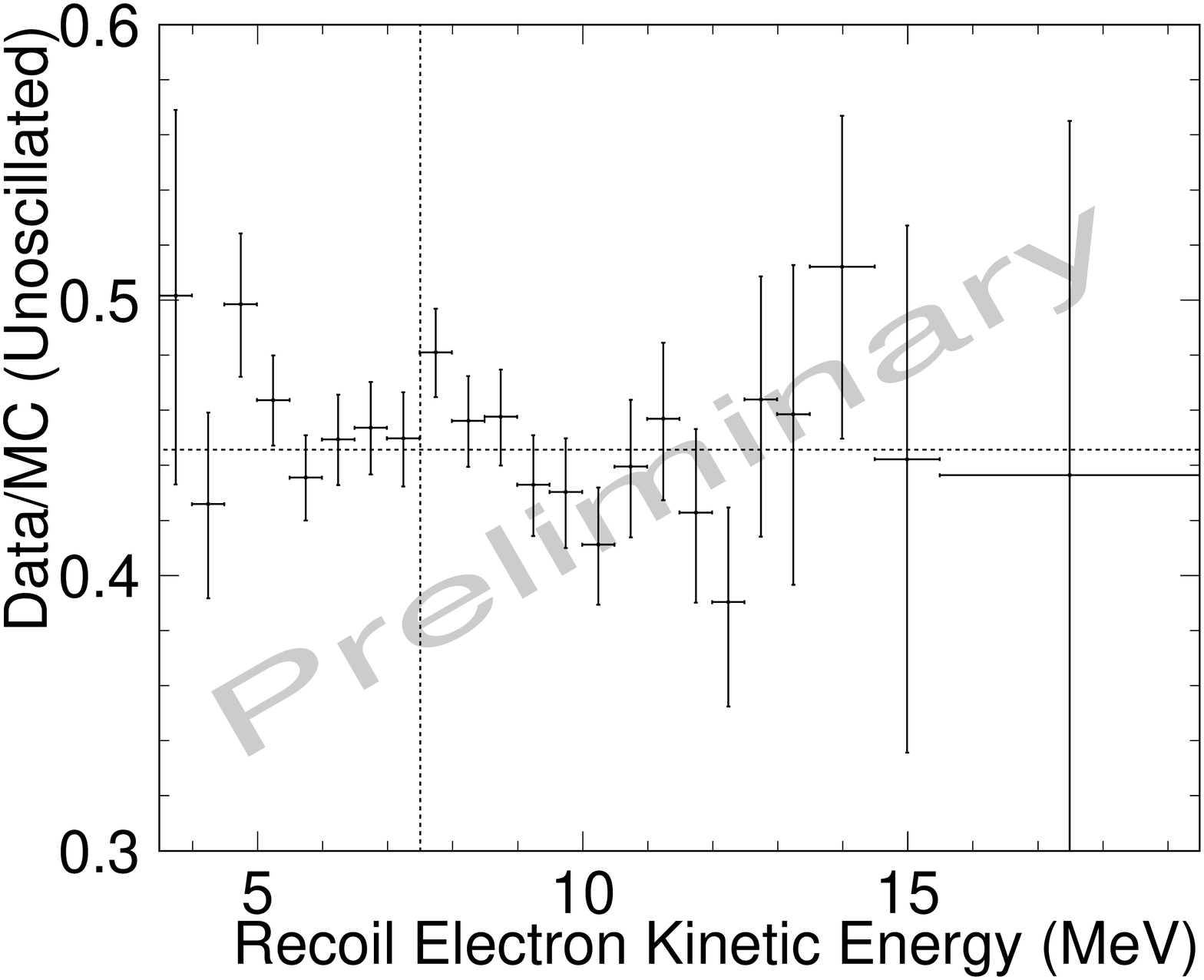}
 \end{center}
\end{subfigure}
\begin{subfigure}{0.5\textwidth}
\vspace*{-0.25cm}
\includegraphics[trim=0cm 0cm 0cm 0.76cm,width=\textwidth,clip=true]{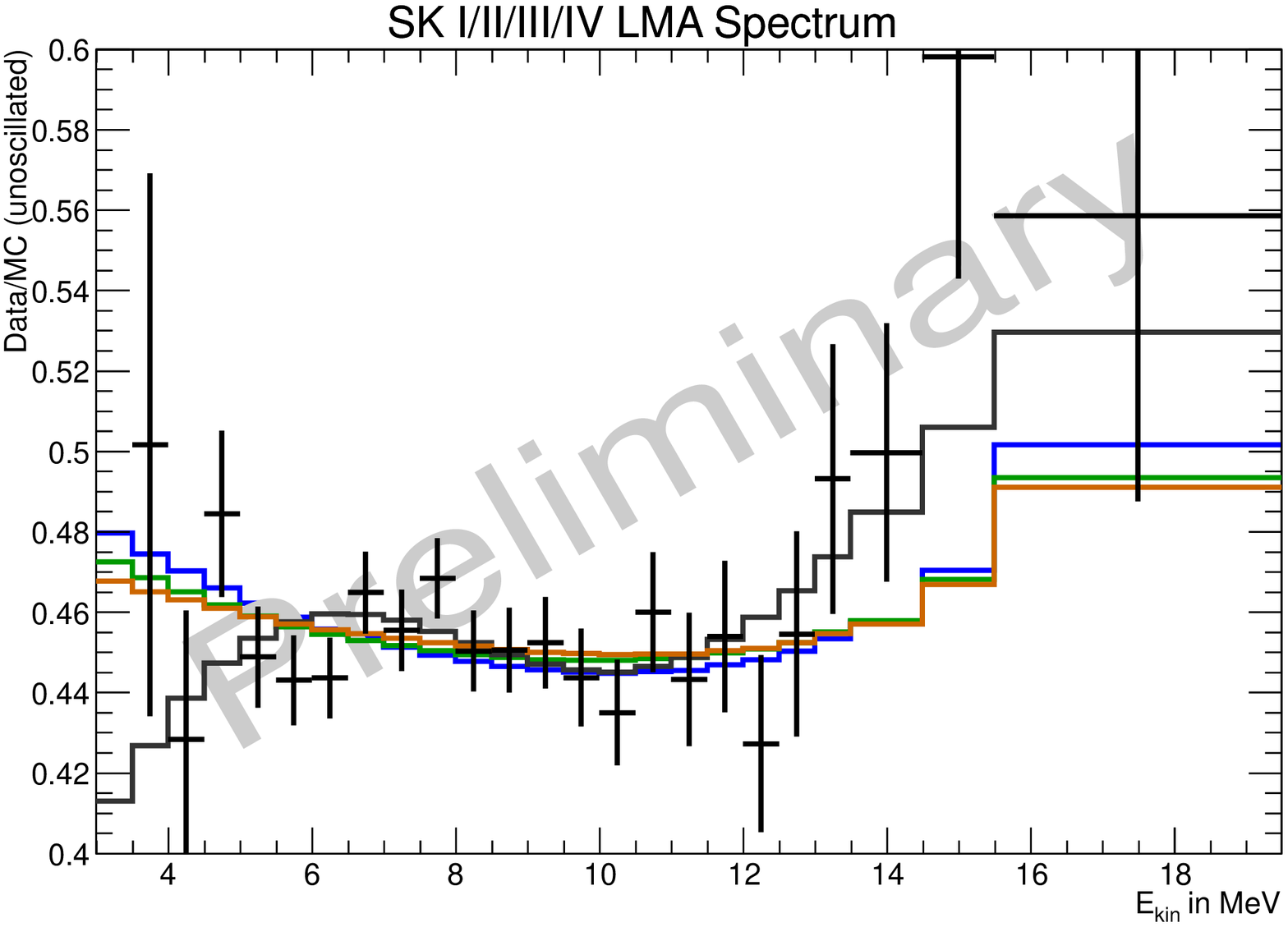}
\end{subfigure}
\caption{Left: SK-IV energy spectrum using MSG sub-samples below 7.5 MeV,
shown as the ratio of the measured rate to the MC simulated unoscillated rate.
 The horizontal dashed line gives the SK-IV total average (0.451).
Error bars shown are statistical plus energy-uncorrelated systematic
uncertainties.  Right: SK-I+II+III+IV recoil
electron spectrum compared to the no-oscillation expectation.
The green (blue) shape is the MSW expectation using the
SK (solar+KamLAND) best-fit oscillation parameters.  The orange (black) line
is the best-fit to SK data with a general exponential/quadratic (cubic)
$P_{ee}$ survival probability.}
\label{fig:spec}
\end{figure}

The SK-IV $^8$B solar neutrino energy spectrum is extracted using the same
method as the total flux, extracting the number of signal events in 23 energy
bins separately.  There are 20 0.5 MeV bins between 3.5 and 13.5 MeV, two 1.0
MeV bins
between 13.5 and 15.5 and one 4.0 MeV energy bin between 15.5 and 19.5 MeV.
Below 7.5 MeV each energy bin is split into three sub-samples based on MSG,
with the boundaries set at MSG=0.35 and 0.45.  The three sub-samples in each
of these low energy bins are simultaneously fit to a single signal and three
independent background components, with the fraction of events in each
sub-sample determined by MC simulated events.  The right panel of
Fig.~\ref{fig:cossun} 
shows the measured angular distributions and fit results for the energy ranges
of 3.5-4.0 MeV, 4.0-4.5 MeV, 4.5-5.0 MeV and 7.0-7.5 MeV.  As expected
in the lowest energy bins, the background component is the largest in the
sub-samples with the lowest MSG, while the signal component grows as the MSG
is increased.  Using this method of MSG sub-samples has reduced the total
uncertainty by up to $15\%$ for the lowest energy bins.
The left panel of Fig.~\ref{fig:spec} shows the resulting SK-IV recoil
electron energy
spectrum, where below 7.5 MeV sub-samples of MSG has been used and above
7.5 MeV the standard signal extraction method is used.  

\subsubsection{SK Combined Solar Neutrino Energy Spectrum Analysis}
\label{combinespec}
\begin{figure}[t]
\begin{subfigure}{0.49\textwidth}
\includegraphics[width=\textwidth]{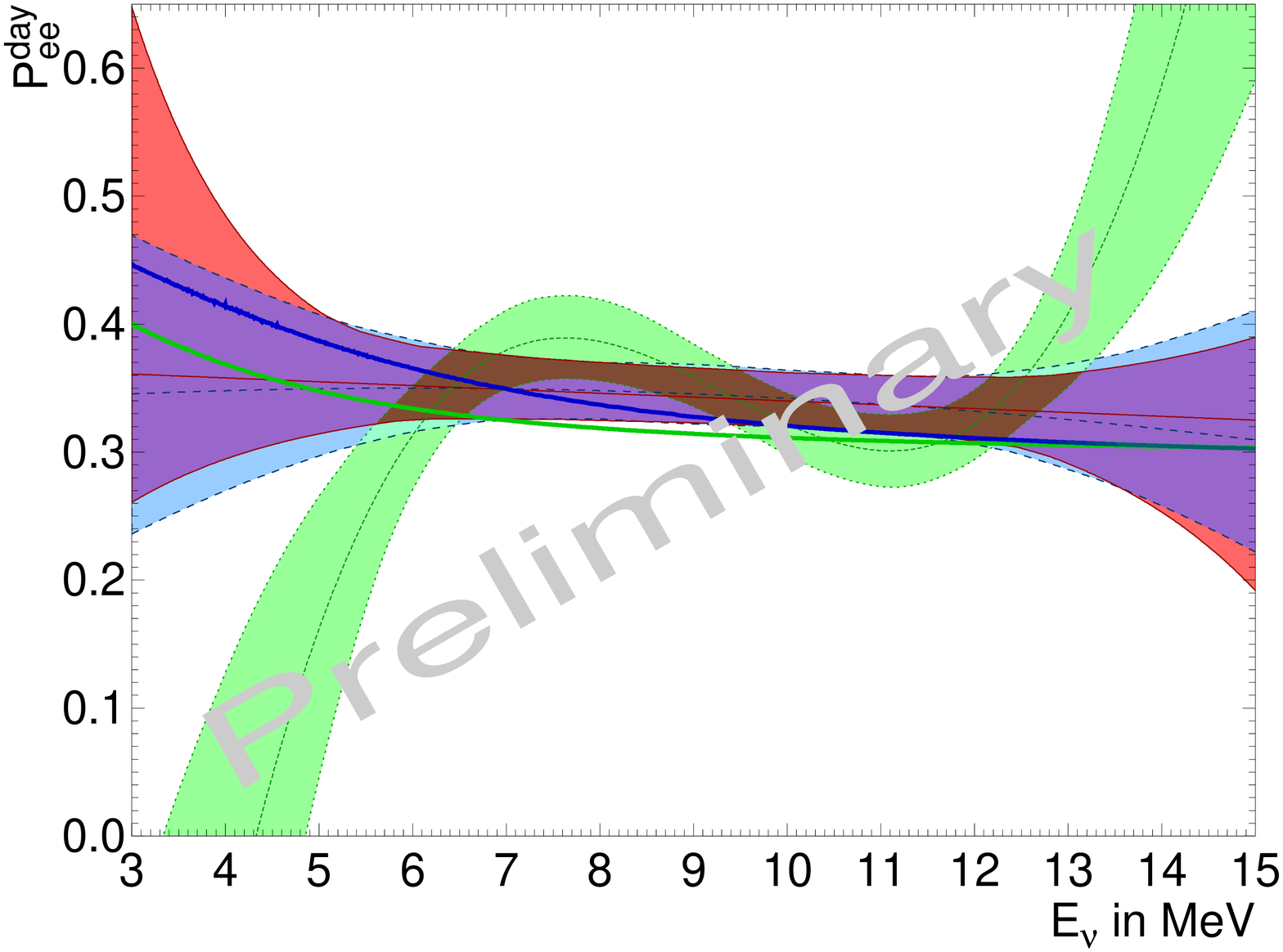}
\end{subfigure}
\begin{subfigure}{0.495\textwidth}
\vspace*{-.25cm}
\includegraphics[width=\textwidth]{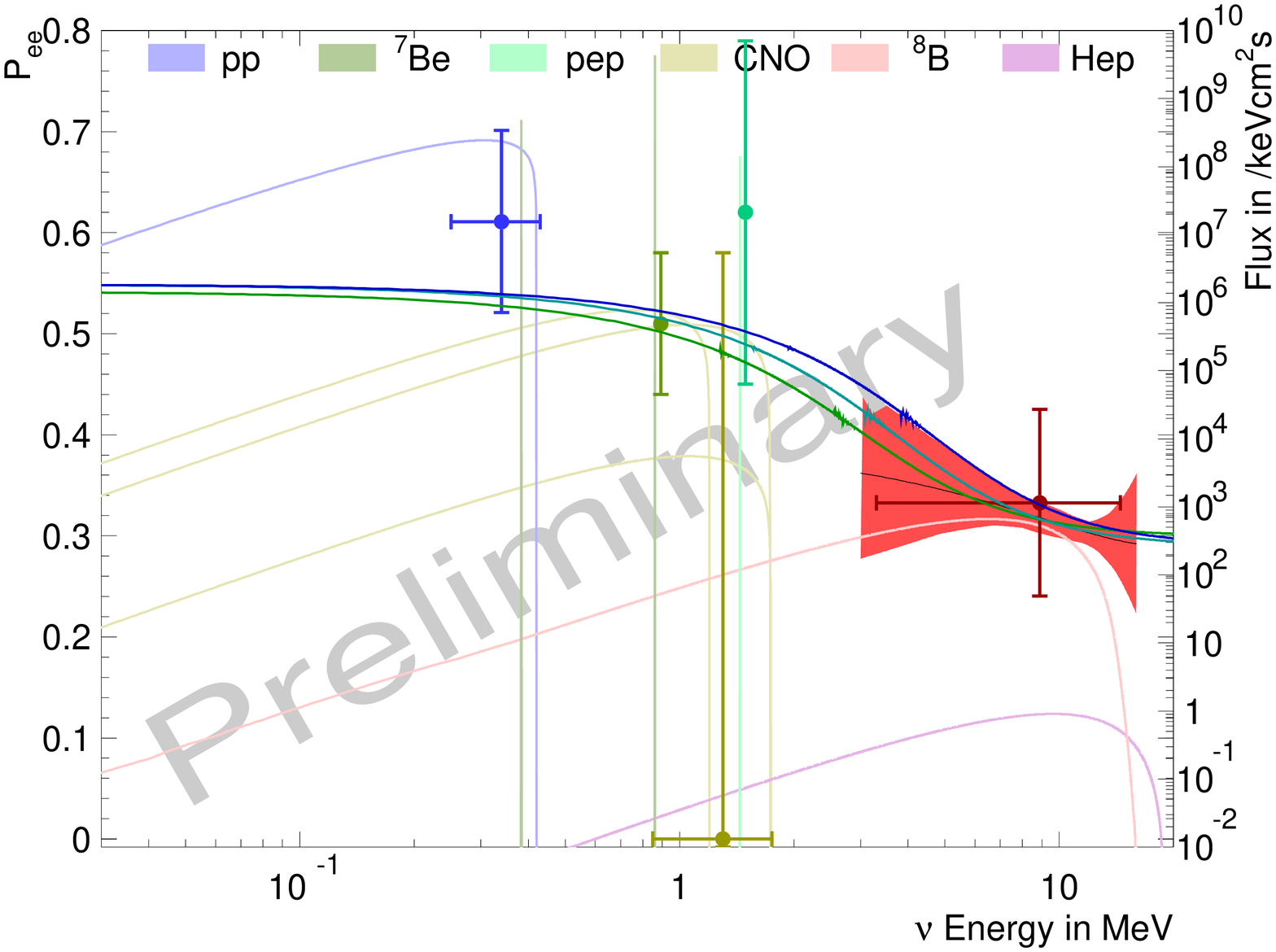}
\end{subfigure}
\caption{Left: Allowed survival probability $1\sigma$ band
from SK data. The red (blue) area is based on an
exponential (quadratic) fit and the green band
is based on a cubic fit. The $^8$B flux is constrained
to the measurement from SNO. The absolute value of the
$^8$B flux doesn't affect the shape constraint much,
just the average value.  Also shown are predictions
based on the oscillation parameters of a fit to all
solar data (green) and a fit to all solar+KamLAND data (blue).
Right: Predicted solar neutrino spectra~\cite{ssm}. Overlaid are expected
MSW survival probabilities; green is the expectation assuming oscillation
parameters from the SK best-fit, turquoise from the global solar 
neutrino best-fit and blue from the solar plus KamLAND best fit.  The
$1\sigma$ band from the combined data of SK and SNO is shown in red.
Also shown are measurements of the $^7$Be (green point),
$pep$ (light green point) and $^8$B flux (red point) by
Borexino~\cite{otherborexino}, as well as $pp$ (blue point) and
CNO values (gold point) extracted from other experiments~\cite{othersolar}.}
\label{fig:pee}
\end{figure}

The spectral data from SK-III has been refit using the same energy bins and
MSG sub-samples as SK-IV, down to 4.0 MeV.  The gain in precision in SK-III
is similar as to SK-IV.  However, in SK-II, the same MSG sub-sample have
been applied for all energy bins.  In order to discuss the energy dependence
of the solar neutrino flavor composition in a general way, the electron
neutrino survival probability $P_{ee}$ has been parameterized using a general
quadratic function $P_{ee}=c_0+c_1(E_{\nu}-10)+c_2(E_{\nu}-10)^2$, as SNO
did in~\cite{snothreephase}, and then by general
exponential and cubic functions as well.  Each phase of SK is fit separately,
and then combined together using a minimum chi-squared method.
The right panel of Fig.~\ref{fig:spec} shows the statistical combination of
the four phases of
SK, along with the best-fits coming from the general quadratic/exponential
(identical and shown in orange) and general cubic (black) function fits.  Also shown in
green (blue) is the expected MSW resonance curves assuming the best-fit
neutrino oscillation parameters coming from a fit to SK data only 
(all solar neutrino plus KamLAND~\cite{kamland} data).  This figure is shown
only as an illustration of the resulting SK combine fit and should not be used
to do further analysis.  Fig.~\ref{fig:pee} shows the resulting $1\sigma$
uncertainties on the spectrum fit to the general functions, along with
the expected MSW curves (same as in Fig.~\ref{fig:spec}).  

There are added benefits when combining the results of the quadratic fit to the 
survival probability of SK and SNO together, since SK's correlation between
the quadratic coefficients $c_1$ and $c_2$ is opposite to that of SNO's.  The
resulting combine $c_1-c_2$ correlation becomes much smaller.  The addition of
the SK data to the SNO data not only significantly increases the precision of
the $c_0$ determination, but the uncertainties on the shape are reduced.  While
SK data by itself prefers an ``upturn'' when going from high to low neutrino
energy and SNO data prefers a ``downturn'', the combined fit favors an
``upturn'' more strongly than the SK data by itself.  SNO's sensitivity is
dominated by charged-current interactions which preserve the neutrino energy,
however, the nuclear threshold energy takes away some of the advantage over
SK, which has higher statistics in the elastic scattering data.  As a consequence, SNO's
uncertainties are smaller at higher neutrino energy, while SK's uncertainties
are smaller at lower neutrino energy.  The right panel of Fig.~\ref{fig:pee}
superimposes the SK plus SNO $1\sigma$ $P_{ee}$ quadratic fit band (red)
(on a logarithmic scale) on the SSM~\cite{ssm} solar neutrino spectrum.  Also
shown are the $pp$ and CNO neutrino flux constraints from all solar neutrino
data~\cite{homestake,othersolar} and the $^7$Be, $pep$ and $^8$B flux
measurements of the Borexino experiment~\cite{otherborexino}.  The SK and SNO
combined allowed band (and the other solar data) are in good agreement with the
predicted MSW curves based on either SK data only, all solar neutrino data or
all solar neutrino plus KamLAND data (shown in green, turquoise and blue,
respectively).

\subsection{Solar Neutrino Day-Night Flux Asymmetry}
\label{dn}
The matter density of the Earth affects solar neutrino oscillations while the
Sun is below the horizon.  This so called ``day-night effect'' will lead to
an enhancement of the $\nu_e$ flavor content during the nighttime
for most oscillation parameters.  The most straight-forward test of this effect
uses the solar zenith angle $\theta_z$ at
the time of each event to separately measure the solar neutrino
flux during the day $\Phi_{\text{\tiny D}}$ (defined as $\cos\theta_z \leq 0$) and
the night $\Phi_{\text{\tiny N}}$ (defined as $\cos\theta_z > 0$). The day-night
asymmetry $A_{\text{\tiny DN}}=(\Phi_{\text{\tiny D}}-\Phi_{\text{\tiny N}})/\frac{1}{2}(\Phi_{\text{\tiny D}}+\Phi_{\text{\tiny N}}$)
defines a convenient measure of the size of the effect.

The SK-IV livetime during the day (night) is 626.4 days (679.9 days).  The
solar neutrino flux between 4.5 and 19.5 MeV and assuming no oscillations
is measured as
$\phi_{\text{\tiny D}}=[2.29\pm0.03(\mbox{stat.})\pm0.05(\mbox{sys.})]\times10^6$ /(cm$^2$sec)
during the day and
$\phi_{\text{\tiny N}}=[2.42\pm0.03(\mbox{stat.})\pm0.05(\mbox{sys.})]\times10^6$ /(cm$^2$sec)
during the night.  By comparing the separately measured day and night fluxes,
the measured day-night asymmetry for SK-IV is found to be
$[-5.3\pm2.0(\mbox{stat.})\pm1.4(\mbox{sys.})]\%$.  When this is combined with
the previous three phases (see the center column of Table~\ref{tab:dn}), SK
measures the day-night asymmetry in this simple
way as $[-4.2\pm1.2(\mbox{stat.})\pm0.8(\mbox{sys.})]\%$~\cite{skall_dn}.
This result deviates from zero by $2.8\sigma$.

\begin{table}[t]
\begin{center}
  \caption{Day-night asymmetry for each SK phase, coming from separate day
and night rate measurements (middle column) and the amplitude fit (right
column). The uncertainties shown are statistical and systematic.
The entire right column assumes the SK best-fit point of oscillation
parameters.}
  \begin{tabular}{l c c}
    \hline\hline
                   &  $A_{\text{\tiny DN}}\pm(\text{stat})\pm(\text{syst})$ & $A_{\text{\tiny DN}}^{\text{\tiny fit}}\pm(\text{stat})\pm(\text{syst})$
    \\ \hline
    SK-I           & $(-2.1\pm2.0\pm1.3)\%$ & $(-2.0\pm1.7\pm1.0)\%$ \\
    SK-II          & $(-5.5\pm4.2\pm3.7)\%$ & $(-4.3\pm3.8\pm1.0)\%$ \\
    SK-III         & $(-5.9\pm3.2\pm1.3)\%$ & $(-4.3\pm2.7\pm0.7)\%$ \\
    SK-IV          & $(-5.3\pm2.0\pm1.4)\%$ & $(-3.4\pm1.8\pm0.6)\%$ \\ \hline
    Combined       & $(-4.2\pm1.2\pm0.8)\%$ & $(-3.2\pm1.1\pm0.5)\%$ \\
    \hline\hline
  \end{tabular}
  \label{tab:dn}
\end{center}
\end{table}

To eliminate systematic effects and increase statistical precision, a more
sophisticated method to test the day-night effect is given
in~\cite{dn,sk1}. For a given set of oscillation parameters, the interaction
rate as a function of the solar zenith angle is predicted. Only the shape of
the calculated solar zenith angle variation is used, the amplitude of it
is scaled by an arbitrary parameter. The extended maximum likelihood fit
to extract the solar neutrino signal is expanded
to allow time-varying signals. The likelihood is then evaluated as a function
of the average signal rates, the background rates and a scaling parameter,
termed the ``day-night amplitude''.  The equivalent day-night asymmetry
is calculated by multiplying the fit scaling parameter with the expected
day-night
asymmetry. In this manner the day-night asymmetry is measured more precisely
statistically and is less vulnerable to some key systematic effects.

Because the amplitude fit depends on the assumed shape of the day-night
variation (given for each energy bin in~\cite{dn} and \cite{sk1}), it
necessarily depends on the oscillation parameters, although with very little
dependence expected on the mixing angles (in or near the large mixing angle
solution and for $\theta_{13}$ values consistent with reactor neutrino
measurements~\cite{reactorexp}).  The fit is run for parameters covering the
MSW region of oscillation parameters
($10^{-9}$ eV$^2\le\Delta{m_{21}^2}\le10^{-3}$ eV$^2$ and $10^{-4}\le\sin^2\theta_{12} < 1$),
for values of $\sin^2\theta_{13}$  between 0.015 and 0.035.  Details of the
estimates of the systematic uncertainties resulting from this method are
given in~\cite{sk4}.  

\begin{figure}[t]
\centering
\begin{subfigure}{0.45\textwidth}
   \includegraphics[width=\textwidth]{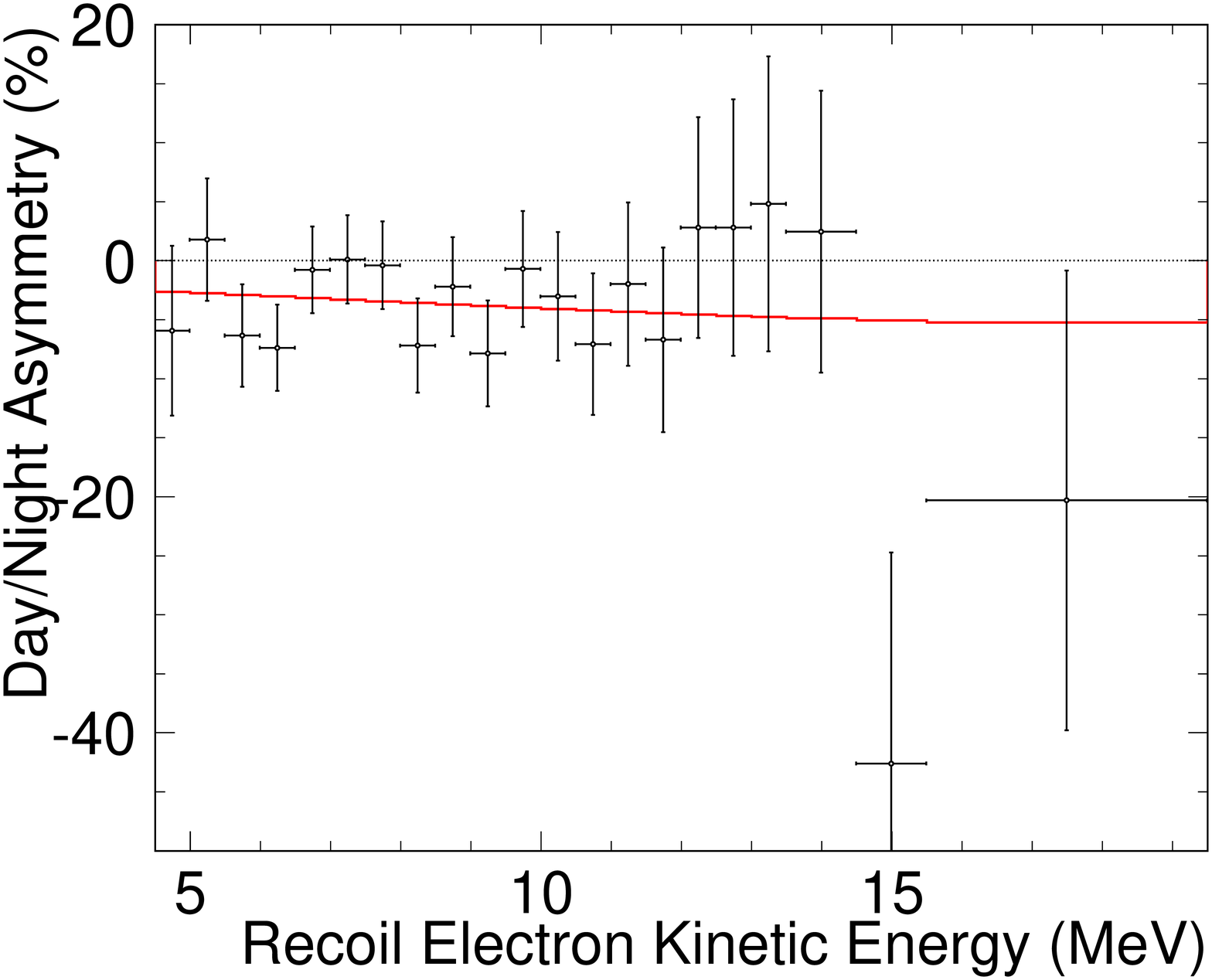}
\end{subfigure}
\begin{subfigure}{0.52\textwidth}
  \centering
    \includegraphics[width=\textwidth]{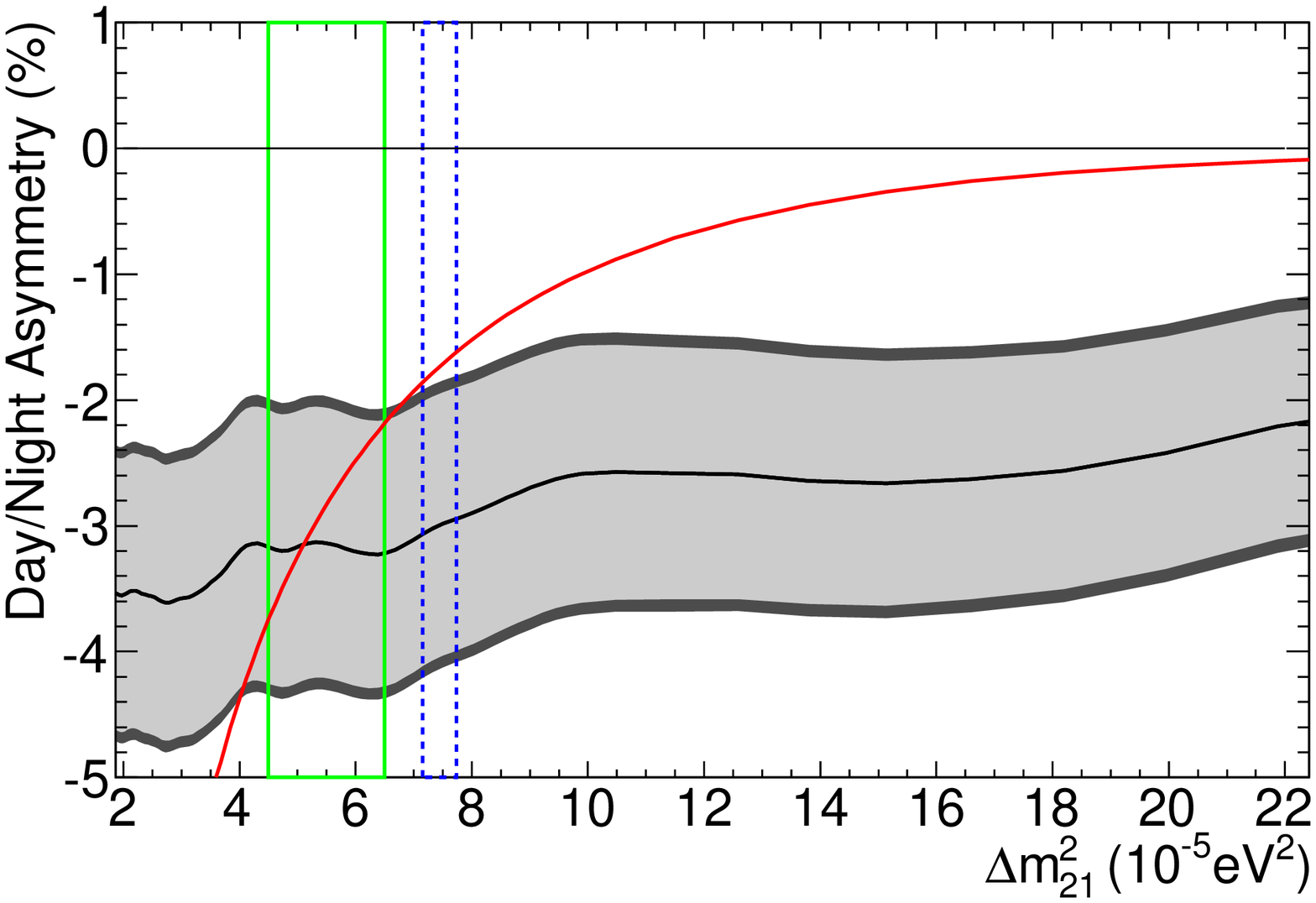}
\end{subfigure}
    \caption{Left: SK day-night amplitude fit as a function of recoil electron
kinetic energy,
shown as the measured amplitude times the expected day-night asymmetry,
for oscillation parameters chosen by the SK best-fit.  The error bars
shown are statistical uncertainties only and the expected dependence is shown
in red.  Right: Dependence of the measured day-night asymmetry
(fitted day-night
amplitude times the expected day-night asymmetry (red)) on $\Delta{m_{21}^2}$,
for $\sin^2\theta_{12}=0.314$ and $\sin^2\theta_{13}=0.025$. The $1\sigma$
stat (stat+syst) uncertainties are given by the light (dark) gray band.
Overlaid are the $1\sigma$ allowed ranges from the solar global fit
(green box) and the KamLAND experiment (blue box).}
    \label{fig:dn}
\end{figure}

The resulting day-night asymmetry when using the extended maximum likelihood
method can be seen for individual phases in the right column of
Table~\ref{tab:dn}.  The left panel of Fig.~\ref{fig:dn} shows the combined
SK-I+II+III+IV day-night amplitude fit as a function of recoil electron
energy.  In each recoil electron energy bin $e$, the day-night variation
is fit to an amplitude $\alpha_e$. The displayed day-night asymmetry
values are the product of the fit amplitude $\alpha_e$ with the expected
day-night asymmetry $A_{\text{\tiny DN, calc}}^e$ (red), when using the SK
best-fit point of oscillation parameters
($\Delta{m_{21}^2}=4.84\times10^{-5}$ eV$^2$,
$\sin^2\theta_{12}=0.342$ and $\sin^2\theta_{13}=0.025$).  These parameters are
chosen when using SK's spectral and time variation data along with constraints
on the $^8$B solar neutrino flux and $\theta_{13}$.
When all energy bins are fit together and the same oscillation
parameters assumed, the resulting SK-measured day-night asymmetry
coming from the amplitude fit is
\begin{align*}
A_{\mbox{\tiny DN}}^{\mbox{\tiny fit}}=[-3.2\pm1.1(\mbox{stat.})\pm0.5(\mbox{sys.})]\%\mbox{  \cite{skall_dn}},
\end{align*}
with an asymmetry of
$-3.3\%$ expected by numerical calculations (see \cite{dn} for details).  This
result deviates from zero by $2.7\sigma$, giving the first significant
direct indication for matter enhanced neutrino oscillations.

If this value is combined with SNO's measurement~\cite{snothreephase}, the
resulting measured SK equivalent day-night asymmetry is
$A_{\mbox{\tiny DN}}^{\mbox{\tiny fit}}=[-2.9\pm1.0(\mbox{stat.+sys.})]\%$, increasing the
significance for a non-zero day-night asymmetry to $2.9\sigma$.  While the
expected day-night asymmetry at SK changes to $-1.7\%$ if the
value of $\Delta m^2_{21}$ is changed to $7.41\times10^{-5}$ eV$^2$ (motivated
by KamLAND data~\cite{kamland}), the measured value is found to be
$A_{\mbox{\tiny DN}}^{\mbox{\tiny fit}}=[-3.0\pm1.0(\mbox{stat.})\pm0.5(\mbox{sys.})]\%$,
reducing the significance for a non-zero day-night asymmetry from 2.7 to
$2.6\sigma$.  The dependence of the SK measured day-night asymmetry on
$\Delta m^2_{21}$, for $\sin^2\theta_{12}=0.314$ and $\sin^2\theta_{13}=0.025$,
can be seen in the right panel of Fig.~\ref{fig:dn}, with the expected
day-night asymmetry
shown by the red curve.  Superimposed are the $1\sigma$ allowed ranges in
$\Delta m^2_{21}$ from the solar global fit~\cite{sk4} (green) and from the
KamLAND experiment~\cite{kamland}.  The resulting day-night asymmetry has
negligible dependence on the values of $\theta_{12}$ (within the LMA region)
and $\theta_{13}$ (near the reactor antineutrino best-fit~\cite{reactorexp}).

\subsection{Solar Neutrino Oscillation Analysis}
\label{osc}

\begin{figure}[t]
\begin{subfigure}{0.5\textwidth}
  \centering
  \includegraphics[width=\textwidth]{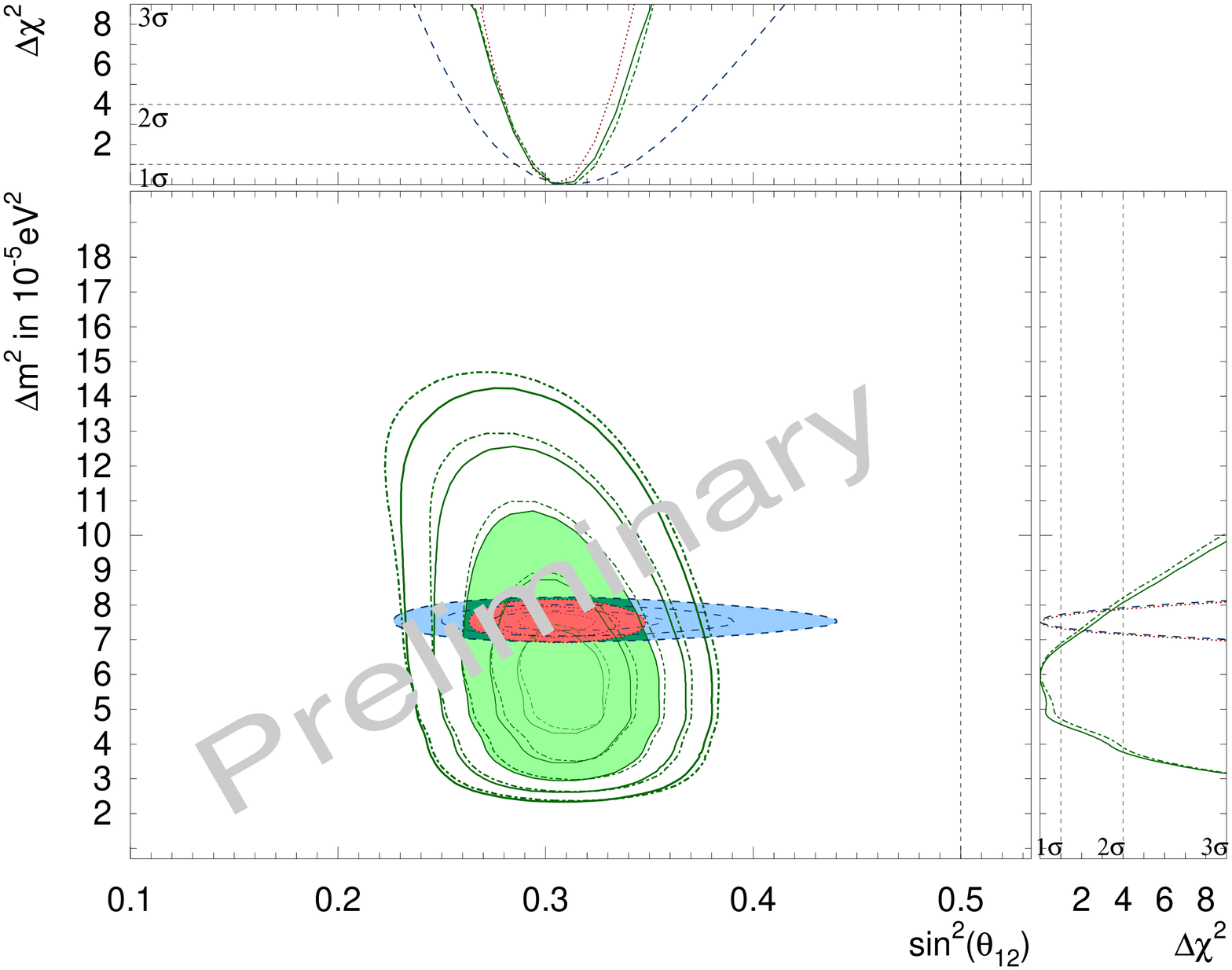}
\end{subfigure}
\begin{subfigure}{0.5\textwidth}
  \centering
  \includegraphics[width=\textwidth]{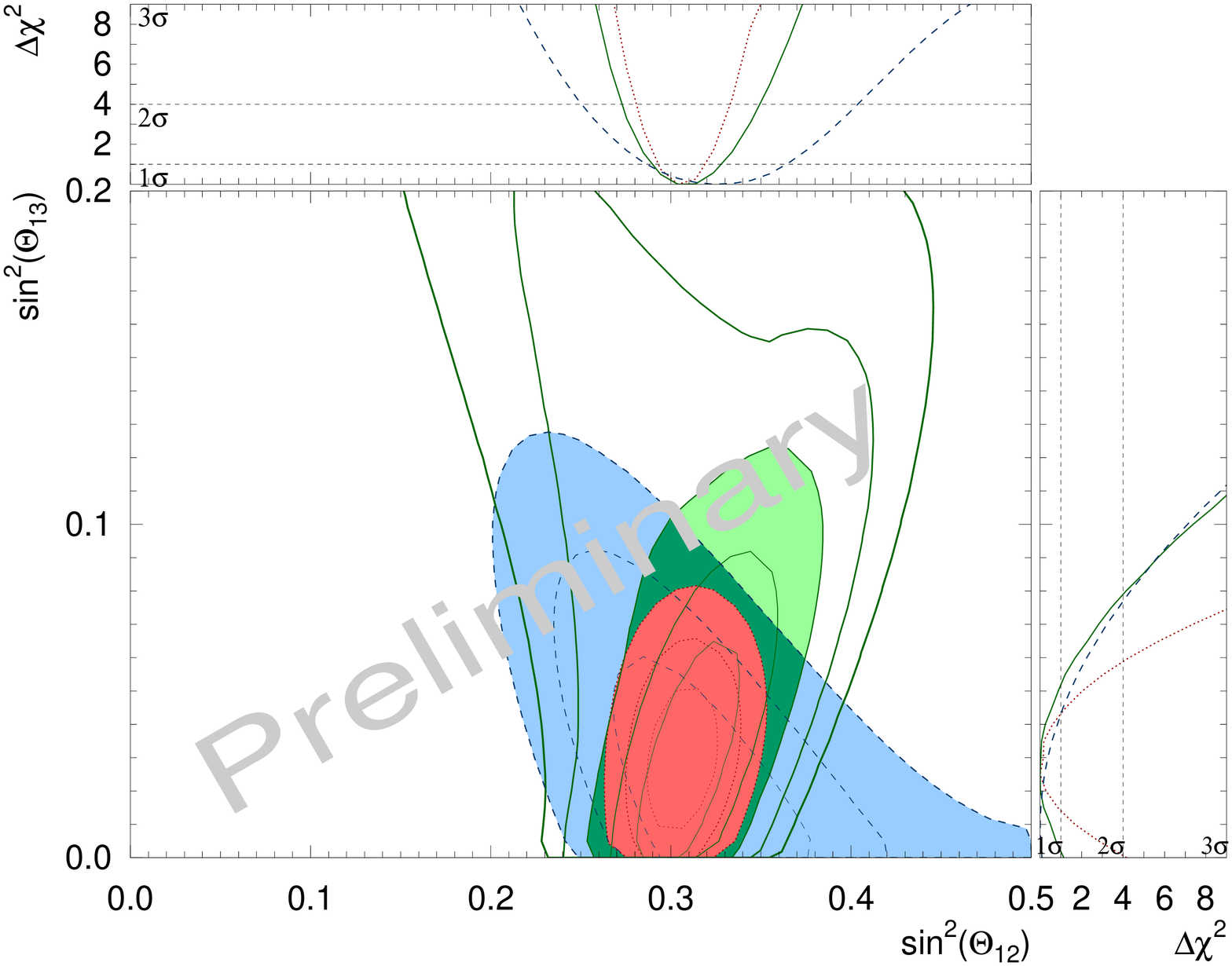}
\end{subfigure}
\caption{Left: Allowed contours of $\Delta m^2_{21}$ vs. $\sin^2\theta_{12}$
from solar neutrino data (green)
at 1, 2, 3, 4 and $5\sigma$ and KamLAND data (blue) at
the 1, 2 and $3\sigma$ confidence levels. Also shown is the
combined result in red. For comparison, the almost identical
result of the SK+SNO combined fit is shown by the dashed
dotted lines. The filled regions give the $3\sigma$ confidence levels.  
$\theta_{13}$ is constrained by 
$\left(\frac{\sin^2\theta_{13}-0.0242}{0.0026}\right)^2$.
Right: Allowed contours of $\sin^2\theta_{13}$ vs.
$\sin^2\theta_{12}$, colors are the same as the left panel.}
  \label{fig:osc}
\end{figure}

We analyzed the SK-IV elastic scattering rate, the recoil electron spectral
shape and the day-night variation to constrain the solar neutrino oscillation
parameters.  We then combined the SK-IV constraints with those of the previous
three SK phases, as well as all other solar neutrino experiments.  The allowed
contours of all solar neutrino data (as well as KamLAND's constraints) are
shown in Fig.~\ref{fig:osc}.
SK and SNO dominate the combined fit to all
solar neutrino data.  This can be seen from the almost identical two sets
of green contours in the left panel of Fig.~\ref{fig:osc}.
In the side panel of this figure, some tension between the
solar neutrino and reactor antineutrino measurements
of the solar $\Delta m^2_{21}$ is evident, stemming from the SK day-night
measurement. Even though the
expected amplitude agrees within $\sim1.1\sigma$ with the fitted
amplitude for any $\Delta m^2_{21}$,
in either the KamLAND or the SK range, the SK data somewhat favor the shape of
the variation predicted by values of $\Delta m^2_{21}$ that are smaller than
KamLAND's. The right panel of Fig.~\ref{fig:osc} shows the results of
the $\theta_{13}$
unconstrained fit.  The significance of non-zero $\theta_{13}$ from the
solar+KamLAND data combined fit is about $2\sigma$, measured as
$\sin^2\theta_{13}=0.026^{+0.017}_{-0.012}$ and quite consistent with
reactor antineutrino measurements~\cite{reactorexp}.

\section{Conclusion}
\label{conclusion}
The fourth phase of SK measured the solar $^8$B neutrino-electron
elastic scattering-rate with the highest precision yet.  When combined with the
results from the previous three phases, the SK combined flux is\\
$[2.37\pm0.015$(stat)$\pm0.04$(syst)$]\times10^6$ /(cm$^{2}$sec).
A quadratic fit of the electron-flavor survival probability as a function of
energy to all SK data, as well as a combined fit with SNO solar neutrino data,
slightly favors the presence of the MSW resonance.
The solar neutrino elastic scattering day-night rate asymmetry is
measured as [$-3.2\pm1.1$(stat)$\pm0.5$(syst)]$\%$.  This solar zenith angle
variation data gives the first significant indication for matter enhanced
neutrino oscillation, and leads SK to having the world's
most precise measurement of $\Delta m_{21}^2=4.8^{+1.8}_{-0.9}$ eV$^2$,
using neutrinos
rather than anti-neutrinos.  There is a slight tension of $1.5\sigma$
between this value and KamLAND's measurement using reactor
anti-neutrinos. The tension increases to $1.6\sigma$, if other solar
neutrino data are included.  A $\theta_{13}$ constrained fit to all solar
neutrino data and KamLAND yields $\sin^2\theta_{12}=0.305\pm0.013$ and
$\Delta m_{21}^2=7.49^{+0.19}_{-0.17}\times10^{-5}$ eV$^2$. When this
constraint is removed, solar neutrino experiments and KamLAND measure
$\sin^2\theta_{13}=0.026^{+0.017}_{-0.012}$, a value in good agreement
with reactor antineutrino measurements.




\bibliographystyle{elsarticle-num}
\bibliography{<your-bib-database>}



\end{document}